# Distances to Local Group Galaxies via Population II, Stellar Distance Indicators. II. The Fornax Dwarf Spheroidal[*]

Elias K. Oakes[1,2], Taylor J. Hoyt[2], Wendy L. Freedman[2,3], Barry F. Madore[2,4], Quang H. Tran[5], William Cerny[2], Rachael L. Beaton[4,6], and Mark Seibert[4]

[1] Department of Physics, University of Connecticut, 196A Auditorium Road, Storrs, CT 06269, USA; elias.oakes@uconn.edu
[2] Department of Astronomy & Astrophysics, University of Chicago, 5640 South Ellis Avenue, Chicago, IL 60637, USA
[3] Kavli Institute for Cosmological Physics, University of Chicago, 5640 South Ellis Avenue, Chicago, IL 60637, USA
[4] The Observatories of the Carnegie Institution for Science, 813 Santa Barbara Street, Pasadena, CA 91101, USA
[5] Department of Astronomy, University of Texas at Austin, 2515 Speedway, Austin, TX 78712, USA
[6] Department of Astrophysical Sciences, Princeton University, 4 Ivy Lane, Princeton, NJ 08544, USA

Received 2021 September 8; revised 2022 March 2; accepted 2022 March 3; published 2022 April 19

## Abstract

We determine three independent Population II distance moduli to the Fornax dwarf spheroidal (dSph) galaxy, using wide-field, ground-based *VI* imaging acquired with the Magellan-Baade telescope at Las Campanas Observatory. After subtracting foreground stars using Gaia EDR3 proper motions, we measure an *I*-band tip of the red giant branch (TRGB) magnitude of $I_0^{\mathrm{TRGB}} = 16.753 \pm 0.03_{\mathrm{stat}} \pm 0.037_{\mathrm{sys}}$ mag, with a calibration based in the LMC giving a distance modulus of $\mu_0^{\mathrm{TRGB}} = 20.80 \pm 0.037_{\mathrm{stat}} \pm 0.057_{\mathrm{sys}}$ mag. We determine an RR Lyrae (RRL) distance from template mean magnitudes, with periods adopted from the literature. Adopting a Gaia DR2 calibration of first overtone RRL period–luminosity and period–Wesenheit relations, we find $\mu_0^{\mathrm{PLZ}} = 20.74 \pm 0.01_{\mathrm{stat}} \pm 0.12_{\mathrm{sys}}$ mag and $\mu_0^{\mathrm{PWZ}} = 20.68 \pm 0.02_{\mathrm{stat}} \pm 0.07_{\mathrm{sys}}$ mag. Finally, we determine a distance from Fornax's horizontal branch (HB) and two galactic globular cluster calibrators, giving $\mu_0^{\mathrm{HB}} = 20.83 \pm 0.03_{\mathrm{stat}} \pm 0.09_{\mathrm{sys}}$ mag. These distances are each derived from homogeneous IMACS photometry, are anchored to independent geometric zero-points, and utilize different classes of stars. We therefore average over independent uncertainties and report the combined distance modulus $\langle\mu_0\rangle = 20.770 \pm 0.042_{\mathrm{stat}} \pm 0.024_{\mathrm{sys}}$ mag (corresponding to a distance of $143 \pm 3$ kpc).

*Unified Astronomy Thesaurus concepts:* Fornax dwarf spheroidal galaxy (548); Population II stars (1284); Distance measure (395); RR Lyrae variable stars (1410); Red giant tip (1371); Horizontal branch (2048); Local Group (929)

*Supporting material:* machine-readable table

## 1. Introduction

The Hubble constant $H_0$ parameterizes the rate of expansion of the universe, with its precise value being foundational to the modern "concordance" cosmological model. Since the HST Key Project established the value of the Hubble constant to within 10% (Freedman et al. 2001), a discrepancy between local $H_0$ measurements from observations of galaxy velocities (Riess et al. 2016) and measurements derived from anisotropies in the cosmic microwave background (Planck Collaboration et al. 2014) has emerged. Taking quoted uncertainties at face value, these two methods now differ by more than $4\sigma$ (Riess et al. 2019; Planck Collaboration et al. 2020). The discrepancy could point to new, not-yet-understood physics, or systematic errors affecting either local or cosmological derivations of $H_0$. For this reason, a more thorough understanding of systematics in the cosmic distance ladder is essential for addressing whether or not (or, to what degree) the $H_0$ tension demands new physics (Freedman 2017).

Traditionally, the distance ladder "rungs" are anchored by geometric parallaxes to stars in the Milky Way. These parallaxes are used to calibrate Cepheid variable period–luminosity (PL) relations, which then serve as calibrators for type Ia supernovae (SNe Ia, e.g., Branch 1998). The Carnegie-Chicago Hubble Program (CCHP) has refined an alternate distance ladder based on the tip of the red giant branch (TRGB), presenting an $H_0$ measurement with an accuracy of 2.7% (Freedman et al. 2019), which has been updated recently incorporating a larger number of calibrating galaxies in Freedman (2021a).

In this and a companion paper, we measure distances to the Fornax (this study) and Sculptor (Q. H. Tran et al. 2022, in preparation) dwarf spheroidal galaxies (dSphs) using the TRGB, RRLs, and the ridgeline of the blue horizontal branch (HB). The proximity and relatively old stellar populations of Local Group dSphs such as Fornax and Sculptor (Tolstoy et al. 2009) make them ideal candidates for Population II distance measurements. At distances of ∼140 kpc for Fornax (Table 4) and ∼84 kpc for Sculptor (Martínez-Vázquez et al. 2015), both galaxies are well positioned for imaging by both ground- and space-based telescopes, since RGB stars can be resolved and measured below the saturation limit in either case. This unique overlap between ground- and space-based observations allows us to explore the relations between standard ground-based *VI* photometric definitions and similar bands on board HST (Appendix D), which is critical to constructing an accurate cosmic distance ladder (see, e.g., Freedman et al. 2019; Riess et al. 2019). Additionally, Fornax and Sculptor are both expected to have negligible internal extinction with a low degree of foreground line-of-sight reddening (Schlegel et al. 1998; Schlafly & Finkbeiner 2011).

---

[*] Based in part on observations made with the NASA/ESA Hubble Space Telescope, obtained at the Space Telescope Science Institute, which is operated by the Association of Universities for Research in Astronomy, Inc. under NASA contract NAS 5–26555. These observations are associated with program No. 13691. Additional observations are credited to the Observatories of the Carnegie Institution of Science for the use of Magellan-Baade IMACS.

Fornax has been shown to host young and intermediate-age stellar populations (Buonanno et al. 1999; Saviane et al. 2000; Pont et al. 2004, among others), along with an expected dominant, older component (Battaglia et al. 2006; Hendricks et al. 2014; Weisz et al. 2014). The chemical enrichment observed in Fornax reflects this complex, extended history, with a metallicity range of $-3.0 < [Fe/H] < -0.5$ dex seen in stars outside its globular clusters (Buonanno et al. 1985; Beauchamp et al. 1995; Saviane et al. 2000; Gullieuszik et al. 2007; Hendricks et al. 2014). The presence of multiple overlapping stellar populations makes the galaxy well suited as a benchmark for comparing independent distance indicators such as the TRGB, RRLs, and the HB. Fornax's extended metallicity distribution, however, has historically led to differing metallicity assumptions for the calibration of period–luminosity relations as well as the luminosity of the HB, complicating respective distance derivations (Table 4 and Figure 9, with discussion in Section 4.3). To reduce systematic errors and measure a high-accuracy distance to Fornax, we (a) target TRGB stars in the metallicity-insensitive *I*-band, (b) use external abundance measurements and a conservative error term to account for metallicity effects on RRL PLZ relations, and (c) adopt a conservative estimate of metallicity effects in HB distance measurements.

The structure of this paper is as follows: Section 2 describes ground-based IMACS and space-based HST observations. The methods used to reduce these images and an explanation of our acquisition and calibration of stellar photometry are also given. Section 3 describes the use of the TRGB as a distance indicator and our algorithmic approach to its measurement, leading to the detection and calibration of a TRGB distance modulus. Section 4 presents our sample of RRL light curves, which we use to construct period–luminosity and period–Wesenheit relations for another distance modulus measurement. In Section 5, we determine a distance using a galactic globular cluster calibration of the HB. We discuss consistency with the published literature and with zero-point calibrations in Section 6.

## 2. Data and Photometry

### 2.1. Observations with IMACS and HST

Ground-based observations of Fornax were acquired with the Inamori-Magellan Areal Camera and Spectrograph (IMACS) mounted on the 6.5 m Magellan-Baade telescope at the Carnegie Institution's Las Campanas Observatory in Chile (Dressler et al. 2011). The telescope's wide field of view enables sampling of large numbers of RGB stars throughout the main body of Fornax and well into its field, despite the galaxy's relatively large angular size of $\sim 27\farcm 5$ along the major axis (Forbes et al. 2008). On the nights of 2014 July 24–25 and September 20–21, 19 epochs over 9 pointings were obtained in the Johnsons–Cousins *V* and *I* filters, using the f/4 camera to produce images with a $15\farcm 46 \times 15\farcm 46$ field of view and $0\farcs 2$ pixel$^{-1}$ plate scale. At each pointing, a series of short (10–30 s) and long (300 s or 600 s) exposures were taken, with seeings ranging from $0\farcs 52$ to $1\farcs 43$. We then applied standard bias and flat-field corrections to individually process each chip. The 68 mosaic images were used for subsequent photometry, each having been stitched together from eight respective chips.

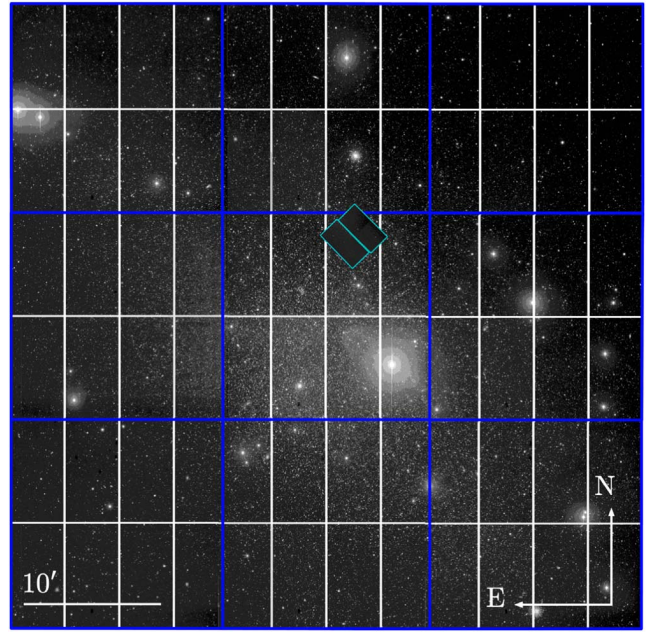

**Figure 1.** Composite mosaic of the Fornax dSph galaxy. Nine IMACS *I*-band pointings covering the full CCHP field are outlined in blue, with two overlapping ACS/WFC F814W pointings outlined in cyan near the field center.

WCS coordinates were added separately to each IMACS pointing by matching images from the Digitized Sky Survey (DSS) to the stitched images and selecting 20–50 Gaia DR2 stars (Gaia Collaboration et al. 2016, 2018) that could be identified in both, while maximizing coverage across the frame. The locations of these stars, in image and Gaia DR2 WCS coordinates, were used to generate a transformation using the AstroPy-associated imexam package and the fit_wcs_from_points() routine (Astropy Collaboration et al. 2013, 2018).

During approximately the same time period, overlapping HST imaging in F606W and F814W was obtained with the ACS/WFC instrument on 2014 September 13–14 (PID: GO-13691, PI: Freedman, Freedman 2014) and retrieved from the Mikulski Archive for Space Telescopes (MAST). The ACS observations consist of two 30 s exposures with a scale equal to $0\farcs 05$ pixel$^{-1}$ and a $202'' \times 202''$ field of view, centered at R. A. = 02:39:47.997, decl. = −34:21:11.78 (J2000). The *flc* images were obtained from the Space Telescope Science Institute (STScI) MAST database. These data products were calibrated, flat-fielded, and CTE-corrected by the STScI pipeline. The *flc* images were also corrected for geometric distortions using ACS pixel area maps. The wide coverage of our ground-based observations and their overlap with deep HST imaging is shown in Figure 1 for selected IMACS *I*-band (blue) and HST F814W (cyan) pointings. These HST observations are used in Appendix D to define transformations from standard ground-based to on-flight photometric systems and intended to bring the ground-based distances presented here onto the ACS on-flight system.

### 2.2. Photometry and Data Reduction

We used the DAOPHOT II and ALLFRAME software suite (Stetson 1987, 1994) to perform empirical point-spread function (PSF) photometry for each IMACS image, following the standard procedure outlined in the DAOPHOT II User Manual. Multiple iterations of PSF subtraction helped to model the brightness profiles of sources near Fornax's crowded globular clusters (Hodge 1961; de Boer & Fraser 2016). We filtered the resulting catalog for high photometric uncertainties and the chi (PSF residual) and sharp (object profile) parameters using a similar approach to Beaton et al. (2019). Finally, we compared long (300–600 s) versus short (10–60 s) exposures in V and I for each image to investigate potential saturation effects for long-exposure photometry near the bright TRGB edge ($I_{\rm IMACS} \approx 8.9$ mag). We found no deviation from linear scatter down to $I_{\rm IMACS} \approx 8.0$ mag; accordingly, all subsequent analysis uses the long-exposure IMACS V and I data.

We reduced the HST imaging separately using the automated CCHP reduction pipeline, following the same procedure except differing by using TinyTim PSF models (Krist et al. 2011) instead of empirically determined ones. A more detailed description of this pipeline is presented in Hatt et al. (2017) and Beaton et al. (2019).

### 2.3. Calibration of IMACS Photometry

We calibrated the IMACS fields by tying in to high-precision photometric standards provided by the Canadian Astronomy Data Centre (CADC, Stetson 2000). Matched sources between the two catalogs were visually inspected to exclude crowded, non-stellar, or otherwise spurious pairs, providing 100-400 calibrators for each image. We investigated the magnitude dependence on $V-I$ color to confirm the absence of any color terms in each calibration (panels (a) and (b) of Figure 2). The weighted flux-averaged mean magnitudes of these pairs, excluding outliers more than three standard deviations from the median, were then used to individually calibrate each corresponding image's photometry (panels (c) and (d) of Figure 2). The matched magnitude difference against Stetson color and magnitude is shown in Figure 2 for a representative pair of 300 s exposures with 270 and 272 matched sources in the V- and I-bands, respectively. Smoothed kernel density estimations (KDEs) for each relation show roughly symmetrical, Gaussian distributions about the mean transformation.

To compute extinction-corrected magnitudes, we applied V- and I-band foreground extinction corrections following Schlafly & Finkbeiner (2011)'s recalibration of the Schlegel et al. (1998) infrared dust map, assuming a Fitzpatrick (1999) reddening law with $R_V = 3.1$. For the adopted extinction corrections (i.e., $A_V = 0.058$ mag and $A_I = 0.032$ mag) we add a systematic error term of half of $A_I$ (0.016 mag) in our final distance moduli to reflect possible reddening underestimates and variation across the field.

After calibrating and applying these reddening corrections, we internally matched duplicate sources across the full IMACS catalog using TOPCAT (Taylor 2005) with a threshold of 0.75″ distance between centroids. Magnitudes of matched sources were then averaged in flux space for a cleaned and calibrated photometric catalog of stars without duplicate objects. We additionally applied a cut to minimize foreground contamination using Gaia EDR3 catalog membership and proper motions for sources near the Fornax center. This process is described in more detail for the Sculptor dSph in Appendix A of our companion paper (Q. H. Tran et al. 2022, in preparation). For Fornax, the Gaia EDR3 proper motion cleaning removed 812 sources.

The resulting $V_0$ (a) and $I_0$ (b) versus $(V-I)_0$ CMDs are shown in Figure 3, where the luminous main sequence meets a wide and well-populated RGB at $I_0 \approx 23$ mag. Past the horizontal branch just above $I_0 \approx 21$ mag, a conspicuous red clump of intermediate-age stars is evident at $I_0 \approx 20.3$ mag. The TRGB discontinuity can be seen at $I_0 \approx 16.8$ mag. We refer the reader to Stetson et al. (1998); Saviane et al. (2000); Pont et al. (2004); Battaglia et al. (2006); and de Boer et al. (2012) for detailed discussions of Fornax's stellar populations.

The calibrated V, I photometric catalog is made available as Data Behind the Figure for Figure 3. The column names and first four lines of data are given in Appendix A. Magnitude errors were estimated as a sum of squares with DAOPHOT magnitude errors and standard deviation of the calibrating image (e.g., Figure 2 panels (c) and (d)), and should not be interpreted as rigorous estimates of the photometric error. Sources excluded by Gaia EDR3 proper motions are flagged with "1" in the CutFlag column. The resulting catalog contains 180,471 sources.

## 3. The Tip of the Red Giant Branch

### 3.1. TRGB as a Distance Indicator

The TRGB is marked by the abrupt truncation of the high-luminosity end of the RGB sequence, as low-mass, hydrogen-burning red giant stars rapidly transition from evolving upwards along the RGB to evolving downwards onto the horizontal branch or red clump (Iben & Renzini 1983; Serenelli et al. 2017). Because the electron-degenerate cores of these stars are unable to modulate temperature through expansion, the onset of helium fusion at the TRGB initiates a runaway thermonuclear reaction that deposits massive amounts of energy in the core as the degeneracy is lifted. This process occurs at an empirically and theoretically well-constrained characteristic luminosity, and thereby serves as an effective Population II distance indicator (Salaris & Cassisi 1997; Rizzi et al. 2007b; Bellazzini 2008).

Since Lee et al. (1993)'s introduction of an algorithmic approach to reproducibly and quantitatively measuring the TRGB, refinements to edge-detection filters and luminosity function (LF) smoothing techniques have enabled modern measurements to better constrain tip detections and quantify their uncertainties (e.g., Madore & Freedman 1995; Hatt et al. 2017; Jang et al. 2018). Furthermore, the TRGB's dependence on color is flattened in the I-band (HST F814W equivalent), where it is insensitive to metallicities as high as $[{\rm Fe/H}] = -0.3$ dex (Barker et al. 2004). By leveraging this transition point in Fornax and in our accompanying analysis of Sculptor, we avoid the need for slope corrections relating to higher-metallicity extensions of the TRGB (Jang & Lee 2017, Q. H. Tran et al. 2022, in preparation).

### 3.2. TRGB Measurement

To quantify the position of the TRGB, we binned the RGB LF at 0.01 mag precision and smoothed its inherent granularity using a Gaussian kernel. We traced a discretely sampled approximation to the first derivative of this smoothed LF by convolving it with a signal-to-noise (S/N) weighted Sobel $[-1, 0, +1]$ edge-detection kernel. The response function is maximized at the dramatic change in local population statistics defined by the TRGB, with false edges effectively suppressed via smoothing.

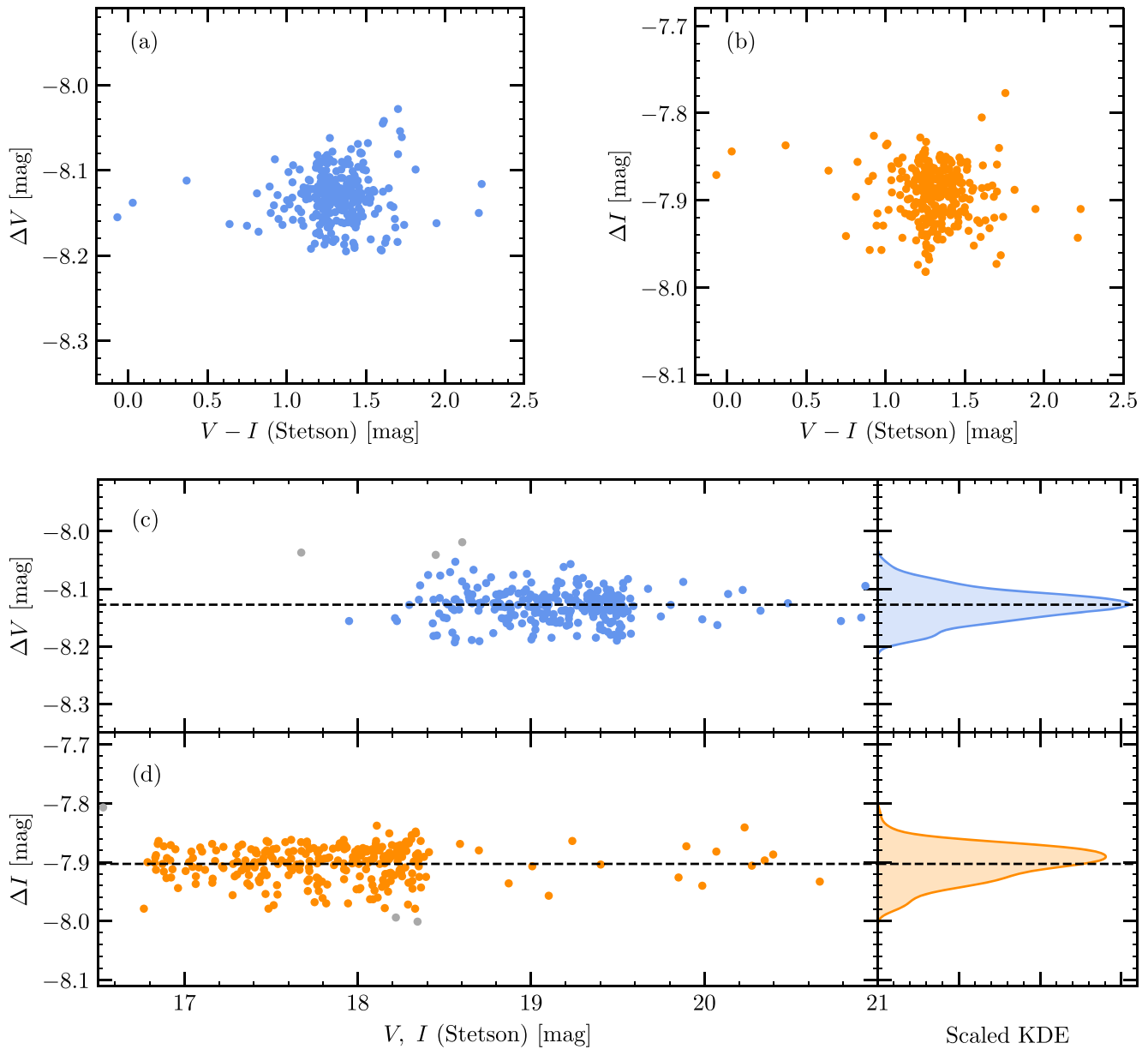

**Figure 2.** Calibration of two (of 68) representative IMACS exposures to high-precision photometry from the CADC: (a) $\Delta V$ (IMACS − Stetson) against Stetson $(V − I)$ color; (b) $\Delta I$ (IMACS − Stetson) against Stetson $(V − I)$ color; (c) $\Delta V$ against $V$ magnitude, with smoothed kernel density estimation (KDE); (d) $\Delta I$ against $I$ magnitude, with smoothed KDE. In panels (c) and (d), weighted flux-space averages are shown as dashed black lines, excluding sources more than $\pm 3\sigma$ from the median (in gray). We find no evidence for a color term or for nonlinearity in this calibration.

As the only free parameter in our smoothing function, the scale $\sigma_s$ must be chosen to minimize LF noise without displacing the TRGB detection. We found that a smoothing scale of $\sigma_s = 0.07$ produced an optimally sharply peaked and symmetrical Gaussian response in the Sobel filter. The left panel of Figure 4 shows the proper motion-cleaned CMD for stars near the TRGB, with boundaries chosen to exclude non-RGB stars assuming an empirical slope of $m_{RGB} = −4.7$ mag color$^{−1}$. Binned (gray) and $\sigma_s = 0.07$ Gaussian-smoothed (red) RGB LFs are shown in the central panel. The right panel, the Sobel kernel response function, reveals an exceptionally well-defined response peak marking our TRGB detection at $I^{TRGB} = 16.785$ mag, or $I_0^{TRGB} = 16.753$ mag after reddening subtraction.

As a test of the edge-detection measurement presented in this section, we also determine a likelihood based measurement of the TRGB. Over the magnitude range $16.25 < I < 18.25$ mag (approximately 0.5 mag above the TRGB and 1.5 mag below it), a broken power law is fit to the Fornax LF (following, e.g., Méndez et al. 2002; Makarov et al. 2006) with four parameters: two slopes $a$ and $b$, TRGB magnitude $m_{TRGB}$, and normalization ratio $c$ at $m_{TRGB}$, along with functions to describe the distribution of photometric errors and completeness. The latter is not considered here because our photometry reaches 6 mag below the TRGB magnitude. It is found that the likelihood minimization is too dependent on the initial guess and adopted

### 3.3. TRGB Error Budget and Calibration

For values of $\sigma_s$ between 0.04 and 0.25 mag, the $I_0^{TRGB} = 16.753$ mag Sobel edge-detection result remains stable within 0.02 mag. This suggests that over-smoothing of the RGB, which should lead to an observable systematic displacement of the kernel peak, does not introduce significant bias at this scale. Nonetheless, we adopt a systematic uncertainty of 0.01 mag to reflect the potential effects of over-smoothing. This value is added in quadrature to conservative estimates of uncertainties due to extinction effects, for which we adopt half the $I$-band foreground extinction (i.e., 0.016 mag) as well as a 0.01 mag term to account for any internal reddening in the galaxy.

Statistical fluctuations in the RGB population can lead to false peaks in the smoothed response function. To estimate the impact of this noise on our detection, we shrink the smoothing parameter $\sigma_s$ until the primary TRGB peak is obscured by noise, finding this occurs at $\sigma_s = 0.03$. We adopt 0.03 mag as the statistical uncertainty on our edge detection. Put another way, we adopt as our statistical uncertainty the width of the response function at the smallest smoothing scale for which the measured peak remains dominant.

We estimate additional statistical uncertainty in our TRGB measurement by restricting the CMD to the color–magnitude range of the RGB population, assuming an empirical slope of $m_{RGB} = -4.7$ mag color$^{-1}$. This cut, applied in the detection of Figure 4, should minimize bias due to asymptotic giant branch stars or foreground contamination. In agreement with CCHP TRGB measurements of other well-resolved samples (e.g., Hatt et al. 2018a, 2018b; Jang et al. 2018), we find that including this cut does not affect our detection and therefore do not increase the error budget.

To investigate the effects of Fornax's population gradient via galactocentric source location, we construct three elliptical annuli using parameters from Battaglia et al. (2006). The radial cuts are defined at $r_{inn} < 10'$, $10' < r_{int} < 19'$, and $19' < r_{out}$ to select for inner, intermediate, and outer stellar populations. This exercise emphasizes a bimodality in Fornax's RGB which can faintly be seen in the full catalog (Figure 4) and becomes more significant at high radii. Fornax's metal-poor population appears, therefore, to be not entirely effectively isolated with a radial cut at $10'$. Along with the dominant tip at $I = 16.785$ mag, this blue RGB feature has a tip at $I = 16.85$ mag, which we detect in the same manner while using a bluer color–magnitude cut along the RGB. The existence of this radial population gradient is well-known in the literature (for instance, it is clearly demonstrated in Figure 6 of de Boer et al. 2012); however, its effect on the TRGB has not been explored. We adopt an additional systematic term of half the difference between these tip populations, i.e., 0.03 mag, to reflect this uncertainty. For the dominant, brighter tip, 0.01 mag variation between our inner and outer samples and our full-sample detection suggests that the influence of a radial population gradient is relatively small and consistent within our uncertainties. This also confirms that internal reddening has only a minor effect on our photometry, which agrees with previous work including McNamara (2011).

The depth and quality of our photometry in Fornax, complemented by its proximity and relatively old stellar population, allow for reasonably accurate uncertainty estimates without simulated artificial star tests. Our photometry reaches significantly deeper than the TRGB with uncertainties fit domain. Thus, we use the *emcee* package (Foreman-Mackey et al. 2013) to better explore the likely parameter space.

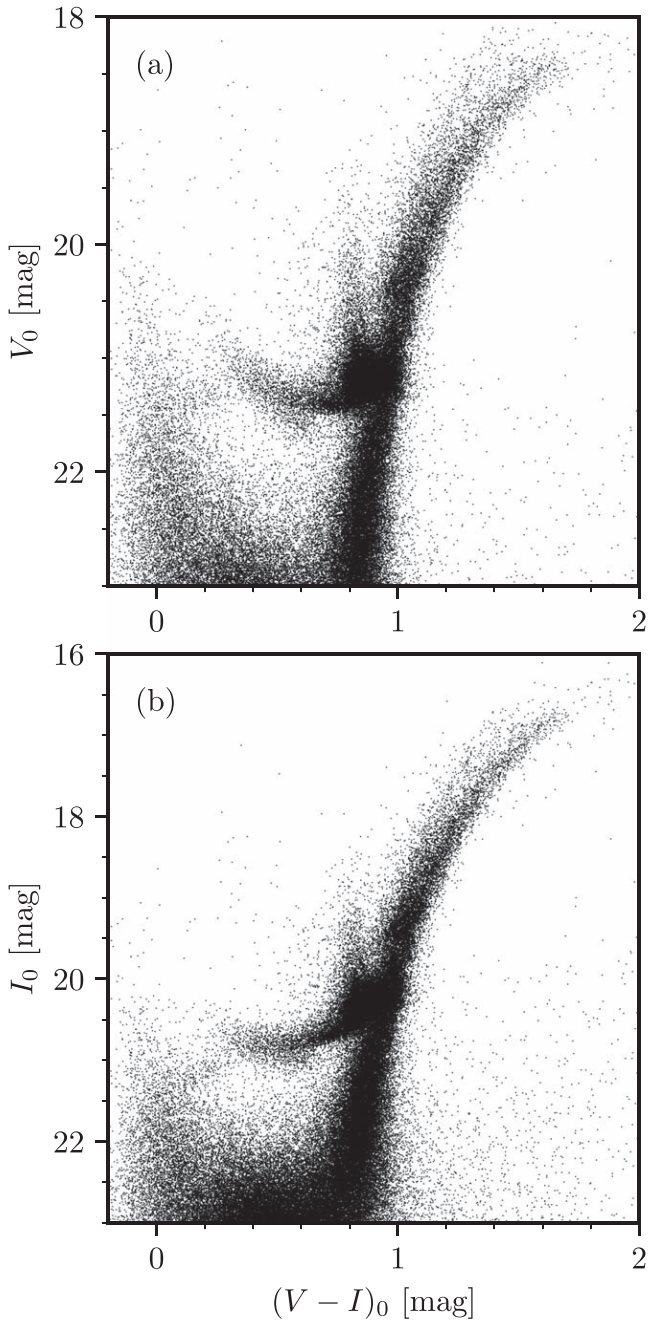

**Figure 3.** Extinction-corrected Fornax dSph $V_0$ (a) and $I_0$ (b) vs. $(V-I)_0$ CMDs using CCHP Magellan-Baade IMACS data. Instrumental magnitudes were transformed to Stetson $V$ and $I$ magnitudes, averaged, and corrected for reddening as described in Section 2.3. 812 bright foreground sources were removed using Gaia EDR3 proper motions.

In Figure 5, the results of the converged MCMC sampling are shown. We find $I^{TRGB} = 16.772^{+0.029}_{-0.027}$ mag, which is in excellent agreement with the edge-detection measurement and uncertainty. Parameter estimates for the RGB slope $a = 0.423^{+0.026}_{-0.025}$ dex mag$^{-1}$, AGB slope $c = 0.462^{+0.380}_{-0.274}$ dex mag$^{-1}$, and logarithmic ratio of RGB to AGB stars at the tip magnitude $b = 0.758^{+0.034}_{-0.034}$ dex are consistent with typical findings (Méndez et al. 2002; Makarov et al. 2006).

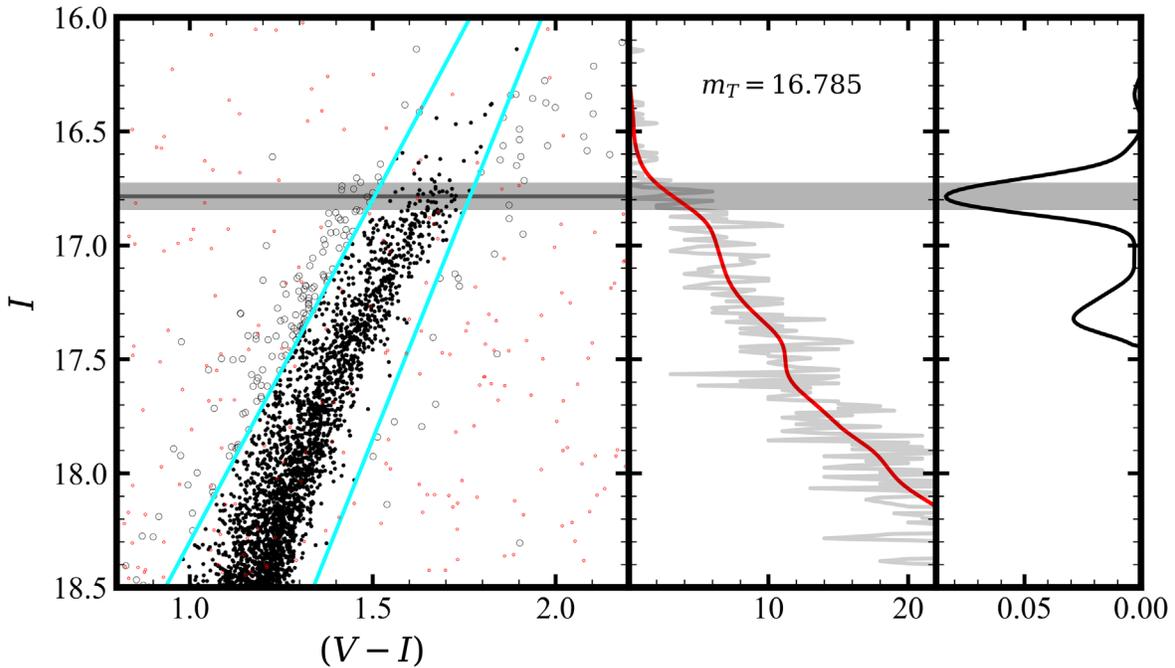

**Figure 4.** Fornax dSph TRGB edge detection (not extinction-corrected). Left: CMD of stars near the TRGB, with the detection at $I = 16.785$ mag as well as the corresponding $2\sigma$ uncertainties (shaded) marked in gray. Red dots show sources removed by Gaia EDR3 proper motion selection, and cyan lines show the color–magnitude selection used to exclude sources shown as open circles. Center: 0.01 mag binned (gray) and $\sigma_s = 0.07$ Gaussian-smoothed (red) luminosity functions. Right: response of the S/N-weighted Sobel $[-1, 0, +1]$ edge-detection kernel on the smoothed LF. The peak response corresponding to the greatest change in the LF shows the TRGB.

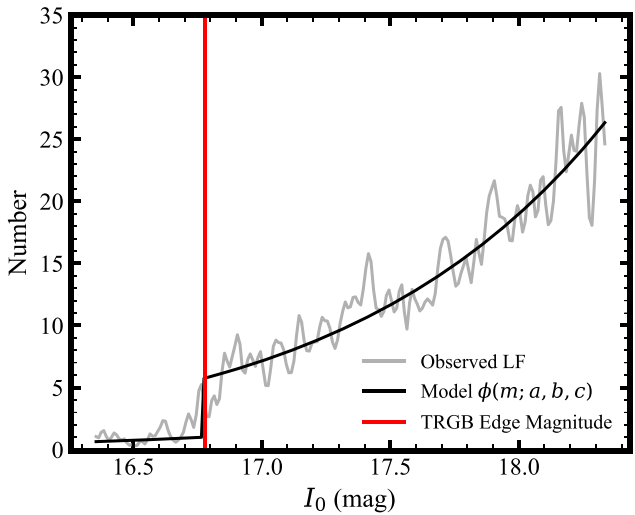

**Figure 5.** Broken power law fit to determine the magnitude of the TRGB. The observed luminosity function (gray) is plotted along with the best-fit AGB+RGB LF (black). For the fit, the model LF is convolved with the photometric error function to better match the behavior of the observed LF.

of $< 0.03$ mag near the tip, indicating a high degree of completeness throughout the RGB. In the well-resolved field of Fornax, we avoid source crowding and systematic effects (Section 2). Furthermore, estimates of Fornax's range of metallicities are consistently well below the threshold for TRGB metallicity sensitivity described by Barker et al. (2004), and an old field population (Battaglia et al. 2006) implies low contamination from young AGB stars. We therefore forego the extensive injection-recovery tests of artificial star LFs used by previous CCHP studies to quantify TRGB detection uncertainties and to optimize the smoothing scale (e.g., Hatt et al. 2017;

Jang et al. 2018; Hatt et al. 2018a, 2018b; Beaton et al. 2019; Hoyt et al. 2019). As summarized in Freedman et al. (2019), these efforts have reliably revealed a statistical edge-detection error of 0.03 mag or less with a similar systematic offset to what we adopt in this work.

Freedman et al. (2020) have provided a calibration of the TRGB anchored to geometric distances of detached eclipsing binaries in the LMC bar. Their quoted zero-point calibration, $M_I^{\mathrm{TRGB}} = -4.047 \pm 0.022_{\mathrm{stat}} \pm 0.039_{\mathrm{sys}}$ mag, includes uncertainty contributions to account for differing photometric zero-points. We additionally incorporate a 0.02 mag systematic term to reflect uncertainty in the photometric zero-point, in our case between Stetson–Landolt and the OGLE calibrators. Freedman (2021a) looked at a wide range of geometric calibrators in different environments, finding zero-points in excellent agreement with this value. With our extinction-corrected TRGB detection of $I_0^{\mathrm{TRGB}} = 16.753 \pm 0.03_{\mathrm{stat}} \pm 0.037_{\mathrm{sys}}$ mag, this calibration returns a TRGB distance modulus to Fornax of $\mu_0^{\mathrm{TRGB}} = 20.80 \pm 0.037_{\mathrm{stat}} \pm 0.057_{\mathrm{sys}}$ mag. Contributions to the TRGB distance modulus error budget are summarized in Section 6.1 and Table 2.

## 4. RR Lyrae Period–Luminosity Relations

### 4.1. RR Lyrae Sample

RR Lyrae variables (RRLs) are old, low-mass, low-metallicity stars with periods of a few hours up to approximately one day. Pulsations are driven by the $\kappa$-mechanism, with stars pulsating in the fundamental mode classified as type a or b and those in the first overtone as type c (Catelan & Smith 2015). As Population II stars whose standard candle nature comes from independent physics to the TRGB, RRLs serve as an excellent test for consistency with our TRGB measurement. Indeed, they are a commonly used distance indicator for the Milky Way and Local Group galaxies, with many studies taking advantage of the known relationship between the mean visual luminosity and mean population metallicity (Demarque et al. 2000). RRLs also exhibit a period–luminosity (PL) relation at longer wavelengths (Longmore et al. 1986; Catelan et al. 2004). To improve calibration of the PL relation, the shorter periods of RRc stars are sometimes "fundamentalized," shifting them by the empirically observed ratio $\Delta \log P = +0.127$ to bring both subtypes on to a common relation (e.g., Braga et al. 2015). Figure 9 and Table 4 include a number of RRL literature distances to Fornax.

The RRL samples used in this study come from Bersier & Wood (2002) and Mackey & Gilmore (2003) (henceforth BW02 and MG03, respectively). Other studies presenting catalogs of variable stars in Fornax were either too sparse (Stringer et al. 2019), incomplete (Clementini et al. 2006; Fiorentino et al. 2017), or targeted Fornax's globular clusters and were therefore affected by crowding in our ground-based photometry (Greco et al. 2009).

BW02 surveyed an area of 0.5 deg$^2$ with 40 and 50 inch ground-based telescopes in the $V$ and $I$ filters, identifying a sample of 515 high-probability RRLs using the Welch–Stetson variability index (Welch & Stetson 1993). Typical photometric errors per epoch for this sample are ~0.15 mag with 20-35 observations for each star, giving well-sampled but noisy light curves. Their characterization of RRL types returned 396 RRab stars and 119 RRc stars, with $\langle P_{ab} \rangle = 0.585$ d and $\langle P_c \rangle = 0.349$ d. Using a period–metallicity relationship, they estimated a mean metallicity of $\langle [\text{Fe/H}] \rangle = -1.64 \pm 0.2$ dex on the Butler-Blanco metallicity scale (Butler 1975), known to be more metal rich than the more commonly used Zinn & West (1984) scale (Bersier & Wood 2002; Rizzi et al. 2007a).

MG03 used archival F555W and F814W HST/WFPC2 imaging to survey four of Fornax's globular clusters, making much of their catalog difficult to match with our IMACS data due to high source crowding. However, we were able to recover a number of uncrowded stars with high-quality light curves at the edges of the globular clusters. They identified 114 RRab and 83 RRc variables using the Welch–Stetson variability of stars near the horizontal branch and fitting light curve templates to estimate periods and mean magnitudes (Layden 1998). Despite having 14-16 high-precision observations for each star, their phased light curves are occasionally undersampled (particularly in the $I$-band) due to observations spanning only a few hours.

We complemented the BW02 and MG03 RRL catalogs by matching stars to our calibrated IMACS photometry, without applying the internal averaging of multi-epoch photometry described in Section 2.3. MG03 published coordinates for their data relative to individual chips instead of a global system, which we converted to WCS coordinates using archival HST data. Overlapping stars were inspected to ensure matches were uncrowded and otherwise uncontaminated, yielding 312 BW02 and 30 MG03 matched RRLs in good alignment and with additional IMACS data.

### 4.2. Light Curves and Phasing

For individual RRLs in the sample, our IMACS photometry typically includes 1–8 epochs in each band, often with an uneven phase-space distribution. The light curves are therefore insufficiently sampled to accurately determine RRL periods or mean magnitudes from this photometry alone. To construct precise PL relations from these sparse data, we additionally utilized the light curves of BW02 and MG03 to determine the amplitude and phase of RRL light curve templates, which were then shifted with a zero-point magnitude offset to match the IMACS data alone. With an appropriate offset term, sampling scaled templates to calculate RRL mean magnitudes helped avoid bias due to sparse light curve sampling from real data.

To assign appropriate light curve templates to the RRLs, it is first necessary to divide them into the two major subtypes which each exhibit different characteristic shapes and periods. The single epoch magnitude uncertainties in BW02's light curves are too high to classify RRL subtypes based only on light curve shapes. We therefore followed BW02 in interpreting the gap in periods at $P \approx 0.46$ days as the transition from RRab to RRc stars, which can be seen in the first column of Figure 7. This divided our full catalog of 342 RRLs into 267 RRab-type stars and 75 RRc-type stars.

The templates we used derive from Beaton et al. (2016), who followed the technique of Freedman (1988) and Freedman & Madore (2010) to generate predictive templates across multiple photometric bands from optical light curves. This method makes use of Gaussian-windowed, Locally Weighted Scatter-plot Smoothing (GLOESS), a nonparametric technique that applies a second-order polynomial fit across a grid of evenly spaced points and weights each observed point with a Gaussian function (Persson et al. 2004; Monson et al. 2017). The requisite data are presented in Monson et al. (2017), who determined GLOESS-smoothed, high-precision optical ($B$, $V$, $I$) light curves of 55 RRLs calibrated to Rich et al. (2018) photometry of 30 galactic RRLs.

Due to the long interval between when the IMACS data and that of BW02 and MG03 were taken, their quoted periods are not precise enough to allow for accurate relative phasing of the IMACS data. To construct a cohesive light curve combining these data sets, we varied the literature periods by a minimum shift to bring the phases into agreement. These shifts were ultimately on the order of $10^{-5}$ days and thus had a negligible impact on the periods.

We used the $V$-band amplitude of these combined literature + IMACS light curves to scale the normalized predictive templates of Beaton et al. (2016). For $I$-band light curves, which exhibit more scatter and are less well sampled, we estimated amplitudes using the ratio $a_I/a_V = 0.65 \pm 0.01$, derived by R. L. Beaton et al. (2022, in preparation) from the light curves of 40 RRab and RRc stars. The phase zero-point for the light curve was defined by maximum light for the $V$-band photometry.

Having scaled each light curve template's amplitude and phase-folded the observed data, we took a two-step approach to fitting for magnitude offsets. First, the template was shifted in both magnitude and phase space relative to the observed light curve to minimize the sum of squared distances between the literature + IMACS data and template. This shift was used to determine the template's zero-point in phase space. With this phasing locked in, the template was shifted only in magnitude space to again minimize the sum of squared distances, this time only considering IMACS data. Crucially, using only IMACS data to calculate the magnitude offset in this second step ensured that our final mean magnitudes do not rely on photometric transformations between literature data and our own, thereby avoiding a potentially significant source of systematic error.

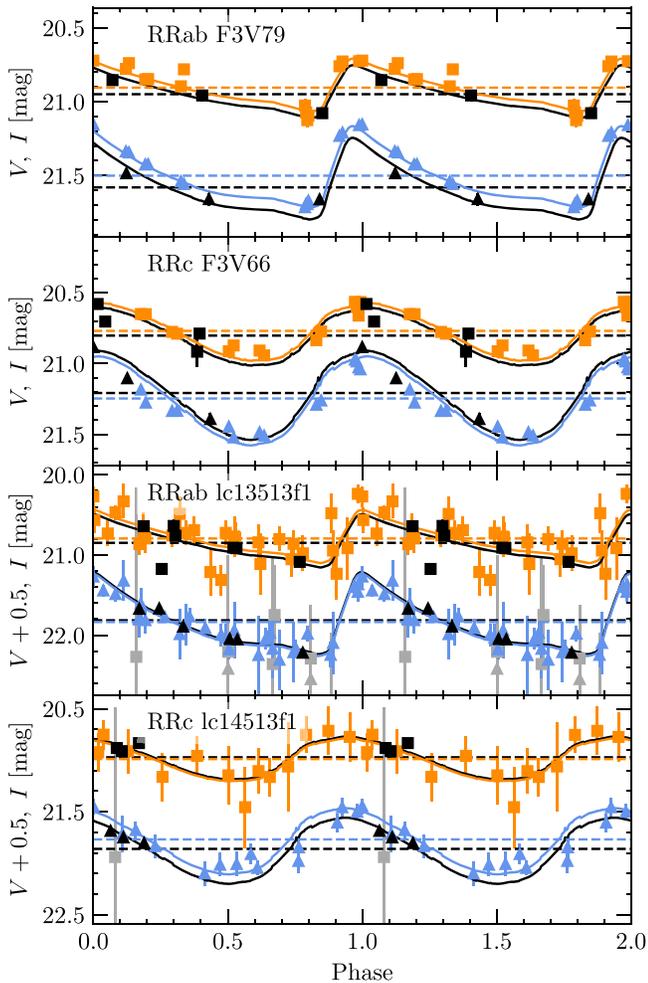

**Figure 6.** Representative *V*- (blue triangles) and *I*-band (orange squares) RR Lyrae light curves from MG03 (top two rows) and BW02 (bottom two rows). Phased IMACS observations are plotted in black, and observations with >0.25 mag (MG03) or >0.5 mag (BW02) per-epoch error, as well as those >2.5$\sigma$ outside the mean magnitudes, were excluded and are shown in gray. Light curve templates were scaled to the combined literature + IMACS data and then offset to match either combined data (colored curves) or IMACS data alone (black curves) as described in Section 4.2. Flux-averaged magnitudes are additionally shown as dashed lines. *V*-band data for BW02's sample are shifted by +0.5 mag for ease of viewing.

In Figure 6, we show representative light curves for two MG03 (top) and two BW02 (bottom) RRLs. The larger amplitudes and asymmetric, saw-toothed light curves typical of RRab-type stars can clearly be seen in contrast to the smaller, sinusoidal light curves of RRc stars. *V*-band literature data are shown as blue triangles, with *I*-band data as orange squares. Phased CCHP IMACS observations are shown in black for both filters. Observations from either data set with a per-epoch error greater than 0.25 mag (MG03) or 0.5 mag (BW02), as well as those outside a threshold of 2.5 times the standard deviation of the light curve magnitudes were excluded and are shown in gray. Interpolated mean magnitudes (dashed horizontal lines) were determined by sampling the scaled, continuous light curve templates (solid curves), and averaging them in flux space.

For the template mean magnitudes, the impact of fitting zero-point magnitude offsets using the full combined light curve data (Figure 6, in color) as opposed to scaling with IMACS data alone (Figure 6, black) was additionally investigated. The amplitude and phase of the IMACS-scaled templates is determined from the combined data set, while the magnitude zero-point remains in the IMACS photometric system. Due to the more limited data in this case, however, this does introduce some additional scatter in derived mean magnitudes. Scatter in mean magnitude derivations was not found to depend on the number of IMACS phase points available. Based on the relatively lower scatter and phase coverage of their light curves, we selected a subsample of 275 RRLs for further analysis. This final catalog of 214 fundamental and 61 first overtone RRLs, with mean magnitudes determined via interpolation of IMACS-scaled light curve templates, is used for all subsequent analysis and period–luminosity detection of the Fornax distance modulus.

### 4.3. Period–Luminosity–Metallicity Relations

Having determined mean magnitudes from IMACS-scaled templates, we apply foreground extinction corrections to the data as described in Section 2.3. In addition to *V* and *I* magnitudes, we consider the associated dual-band Wesenheit magnitude, which takes advantage of the total-to-selective absorption of multiple passbands to minimize the luminosity dependence on reddening and its corresponding uncertainty. Madore (1982) defines this as:

$$W_{I,V-I} = M_I - \xi \cdot (M_V - M_I) \quad (1)$$

where $\xi$ describes the color absorption and excess ratio for $(V-I)$. We adopt the color coefficient $\xi = 1.467$ from Neeley et al. (2019) (henceforth Ne19), their Table 3.

We then fit periods and extinction-corrected $I_0$ and $W_{I,V-I}$ mean magnitudes to a period–luminosity–metallicity relation, adopting Ne19's form:

$$m = a + b \cdot (\log P + 0.30) + c \cdot ([\text{Fe/H}] + 1.36) \quad (2)$$

Due to the extended, tripartite metallicity distribution of Fornax's stellar population, estimating a typical value for our RRL sample is challenging. BW02 determined the metallicity of their RRL catalog via Sandage (1993) period–metallicity relations, arriving at $\langle[\text{Fe/H}]\rangle = -1.64$ dex on the metal-rich Butler-Blanco scale.

We compare this value with high-resolution spectroscopic surveys of Fornax's RGB, starting with Battaglia et al. (2006). Using VLT/FLAMES CaT spectroscopy, they surveyed 562 RGB stars and isolated a metal-poor, older population extending from $-2.5 < [\text{Fe/H}] < -1.0$ dex and a mean $\langle[\text{Fe/H}]\rangle = -1.7$ dex. Hendricks et al. (2014) similarly identified three subgroups in their sample of 340 RGB stars in the outer regions of Fornax, surveyed with the same instrument. They identified metallicity groups centered at $[\text{Fe/H}] = -1.0, -1.4$, and $-1.9$ dex. Notably, they found that the metallicity distributions of Fornax (after removing its most metal-rich population) and Sculptor both peak at $[\text{Fe/H}] = -1.7$ dex, which we adopt for our PLZ fits. We likewise consider period–luminosity and period–Wesenheit relations without a metallicity term.

In order to avoid added dispersion from the RRc PL relation, which exhibits intrinsically higher scatter than the RRab PL relation, we use only the first overtone (RRa- and RRb-type) stars for our primary analysis. This limits our RRL sample to 214 stars. Both populations, as well as a fundamentalized relation for reference, are presented in Figure 7.

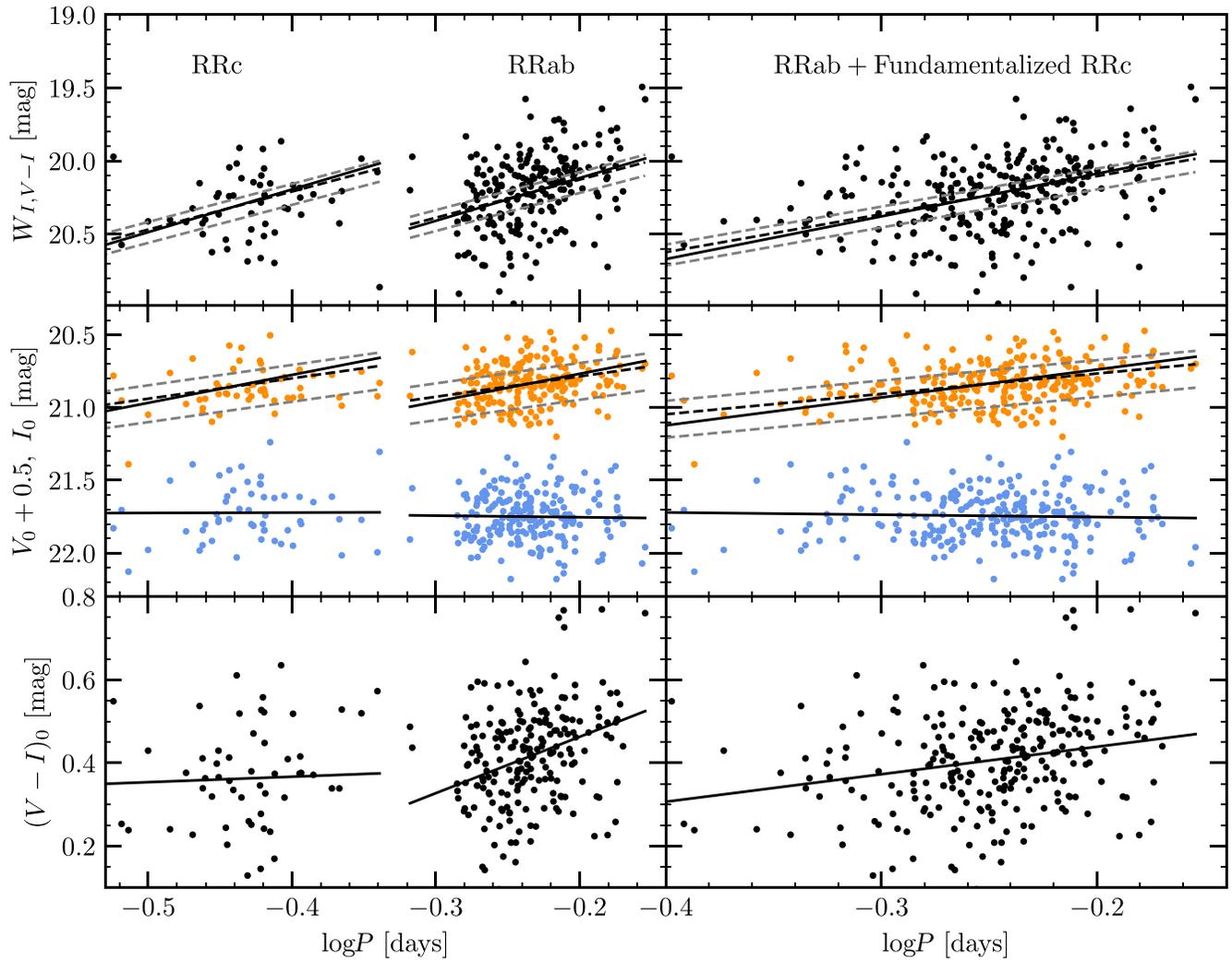

**Figure 7.** Left: extinction-corrected period–Wesenheit (top), period–luminosity (center), and period-color (bottom) relations for 275 Fornax RR Lyraes using the BW02 and MG03 catalogs. Periods and mean magnitudes were determined from Ne19 light curve templates scaled to IMACS data as described in Section 4.2. We shifted the $V_0$–band observations by +0.5 mag for ease of viewing. Solid lines show best-fit PL and PW relations, while dashed lines show the best-fit PLZ and PWZ calibrations for the $I_0$–band and $W_{I,V-I}$ relations. Dashed gray lines show the full range of metallicity considered ([Fe/H] = −2.0 to −1.0 dex). For the $V_0$–band and $(V-I)_0$ color plots, solid lines show linear fits to the data. Right: the same as the left column, with RRc periods fundamentalized by adding $\Delta \log P = +0.127$. Fit parameters and rms values for the RRab sample are given in Table 1 and demonstrate good agreement with theoretical and empirical calibrations.

**Table 1**
RR Lyrae (ab-type) Period–Luminosity Fit Parameters

| Band | a | b[a] | c[a] | [Fe/H][b] | rms$_{fit}$ | rms$_{theory}$[c] |
|---|---|---|---|---|---|---|
| | | | *PL and PW relations* | | | |
| $I_0$ | 20.83 ± 0.01 | −1.92 ± 0.41 | … | … | 0.14 | 0.14 |
| $W_{I,V-I}$ | 20.22 ± 0.02 | −2.92 ± 0.30 | … | … | 0.25 | 0.25 |
| | | | *PLZ and PWZ relations* | | | |
| $I_0$ | 20.91 ± 0.01 | −1.40 ± 0.30 | 0.23 ± 0.04 | −1.7 | 0.16 | 0.16 |
| $W_{I,V-I}$ | 20.26 ± 0.02 | −2.60 ± 0.25 | 0.13 ± 0.03 | −1.7 | 0.25 | 0.25 |

**Notes.**
[a] Neeley et al. (2019).
[b] Discussion in Section 4.3.
[c] Marconi et al. (2015).

The best-fit parameters for the PL relation fits are also plotted in Figure 7 and enumerated in Table 1, using Ne19's empirically derived slopes calibrated to Monson et al. (2017) HST and Gaia geometric parallaxes of galactic RRL stars. For the *I*-band luminosity and Wesenheit relations, solid black lines show linear PL and PW fits while dashed lines show PLZ and PWZ fits. We additionally compare rms values for these fits with theoretically derived slopes from hydrodynamical pulsation models (Marconi et al. 2015) in Table 1. The reported uncertainties on the fit parameters are statistical in nature, arising from the least-squares fit. We find good agreement for both the $I_0$ PL relation and $I, V - I$ PW relation, with the same rms for both the theoretical and empirical fits with and without considering metallicity. This suggests that our data are well-described by the predictions of Marconi et al. (2015).

### 4.4. RRL Distances and Error Budget

Ne19s calibrated relations (their Table 3) offer PL, PW, PLZ, and PWZ zero-point offsets for the RRL distance moduli. Combining these (and their reported uncertainties) with the fit parameters of Table 1, estimated at $\langle \log P \rangle = -0.23$ for our own data, we find distance moduli of $\mu_0^{PL} = 20.66 \pm 0.01_{stat} \pm 0.03_{syst}$, $\mu_0^{PW} = 20.64 \pm 0.02_{stat} \pm 0.03_{syst}$, $\mu_0^{PLZ} = 20.74 \pm 0.01_{stat} \pm 0.03_{syst}$, and $\mu_0^{PWZ} = 20.67 \pm 0.02_{stat} \pm 0.03_{syst}$ mag.

A stricter cut to exclude ambiguously classified and less well-resolved RRLs yields a sample of 59 stars. Using this catalog with IMACS-scaled template mean magnitudes, we find distance moduli consistent to within 0.02 mag. We also consider the effect of fundamentalizing the RRc sample to create a common PL relation for both subtypes, as in the right panel of Figure 7. Fitting the full sample of 275 stars returns $\mu_0^{PLZ} = 20.75$ and $\mu_0^{PWZ} = 20.70$ mag.

We further compare the extensive range of metallicity sampled by Fornax's stellar populations, which covers roughly $-2.0$ to $-1.0$ dex. We consider the full range of these metallicities and their effect on the RRL PLZ and PWZ relations, finding a systematic error of half the distance modulus range equal to 0.11 and 0.06 mag, respectively. This range is represented in Figure 7 by the gray, dashed lines. An uncertainty of 0.01 mag is also adopted to account for potential zero-point differences between the IMACS photometry and the Ne19 photometry. Due to consistent calibration between our data sets and the Stetson CADC standards, this value is likely conservative. Finally, we include the uncertainties of 0.016 mag estimated from our line-of-sight reddening and 0.01 mag internal reddening. For the reddening-free Wesenheit magnitudes, we instead adopt a 0.01 mag uncertainty to reflect potential deviations from the adopted mean Galactic extinction law ($R_V = 3.1$).

These uncertainties are added in quadrature to define the total systematic error, which is summarized along with the statistical error in Section 6.1 and Table 2. We therefore report final RR Lyrae distance moduli of $\mu_0^{PLZ} = 20.74 \pm 0.01_{stat} \pm 0.12_{sys}$ mag and $\mu_0^{PWZ} = 20.69 \pm 0.02_{stat} \pm 0.07_{sys}$ mag, which are consistent with one another and with the TRGB distance modulus.

## 5. Horizontal Branch Fitting

### 5.1. Globular Clusters and HB Calibration

As an additional consistency check on the distances derived through our TRGB detection and RRL PLZ and PWZ relations, we independently establish a distance modulus to Fornax using its blue horizontal branch (HB). Our method involves matching the "empirical isochrone" defined by the HB of a nearby globular cluster (GC) with a well-constrained distance to the CMD of a given dwarf galaxy. This circumvents the use of theoretical isochrones in favor of using the cluster HB ridgelines, which are very well defined and can be calibrated observationally (e.g., Clem et al. 2008).

**Table 2**
Systematic and Statistical Uncertainties in Fornax Distance Moduli

| Value [mag] | Type[a] | Source |
|---|---|---|
| *Tip of the Red Giant Branch* | | |
| 0.03 | Stat. | Minimum smoothing scale |
| 0.022 | Stat. | Freedman et al. (2020) |
| 0.01 | Syst. | Over-smoothing effects |
| 0.03 | Syst. | Secondary tip population |
| 0.039 | Syst. | Freedman et al. (2020) |
| 0.02 | Syst. | Photometric offset |
| 0.016 | Syst. | Foreground extinction |
| 0.01 | Syst. | Internal reddening |
| *RR Lyrae PLZ/PWZ Relation* | | |
| 0.01/0.02[b] | Stat. | Least squares fit |
| 0.11/0.07[b] | Syst. | Metallicity effects |
| 0.01 | Syst. | Photometric offset |
| 0.03/0.02[b] | Syst. | Neeley et al. (2019) |
| 0.016[c] | Syst. | Foreground extinction |
| 0.01[c] | Syst. | Internal reddening |
| 0.01[d] | Syst. | Extinction law |
| *Horizontal Branch Fitting* | | |
| 0.03 | Stat. | Range of GC calibrations |
| 0.05 | Syst. | Color range bias |
| 0.05 | Syst. | DEB distance uncertainty |
| 0.01 | Syst. | Photometric offset |
| 0.04 | Syst. | Intrinsic HB scatter |
| 0.02 | Syst. | GC reddening |
| 0.016 | Syst. | Foreground extinction |
| 0.01 | Syst. | Internal reddening |

**Notes.**
[a] Statistical (Stat.) or Systematic (Syst.).
[b] PLZ/PWZ.
[c] PLZ only.
[d] PWZ only.

To apply such an approach to Fornax, we utilized two GCs with well-constrained distances derived via detached eclipsing binaries (DEBs). These clusters are M55 (NGC 6809) and Omega Centauri ($\omega$ Cen; NGC 5139), which have [Fe/H] metallicities of $-1.94$ dex and $-1.53$ dex, respectively (Harris 1996, 2010 edition). As discussed in Section 4.3, this range approximates the spread in metallicity that describes Fornax's intermediate- and old-age stellar population, including the RR Lyraes. For each of these clusters, we began by selecting member stars using their Gaia EDR3 proper motions, following the method described in Cerny et al. (2020). We then cross-matched the resulting catalog with the homogeneous cluster photometry provided by Stetson et al. (2019), which is calibrated onto the same photometric system (Landolt standards) as our Fornax IMACS photometry. We defined an empirical ridgeline fiducial by sampling the median magnitude of HB stars across the range of photometric colors that encompasses each cluster's blue HB and fitting a smoothed spline to these data points.

We calibrate these cluster fiducials by adopting the following DEB distances and reddening values: For M55, we adopt a distance modulus of $\mu_0 = 13.58 \pm 0.05$ mag with reddening $E(B - V) = 0.115$ mag (Kaluzny et al. 2014). For $\omega$ Cen, we take the revised uncertainty discussed in Cerny et al. (2020) and adopt $\mu_0 = 13.68 \pm 0.10$ mag, with reddening $E(B - V) = 0.12$ mag (Thompson et al. 2001). The empirically defined HB fiducials for M55 and $\omega$ Cen are scaled by the appropriate zero-point distance modulus and shown in Appendix B.

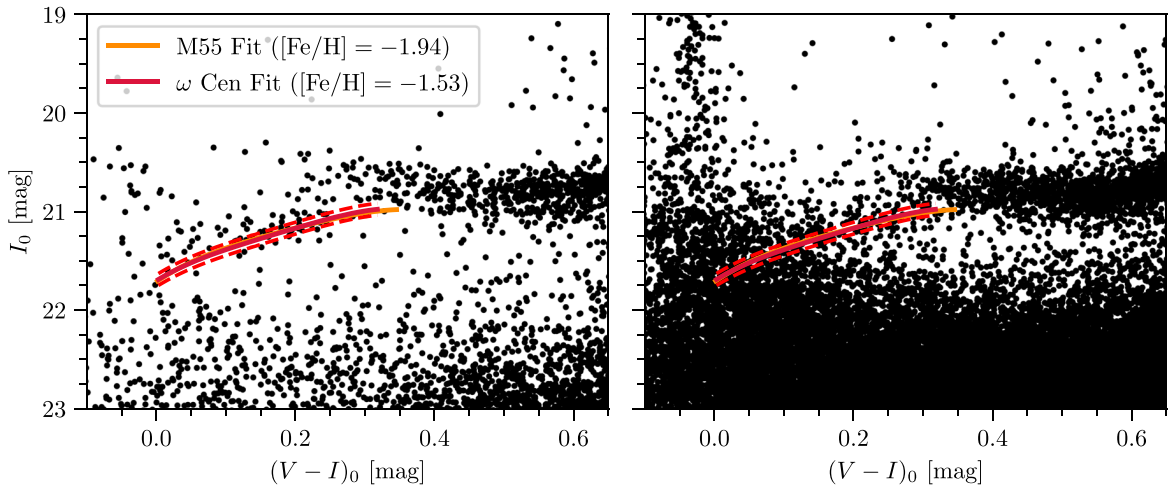

**Figure 8.** Left: Fornax's extinction-corrected CMD centered on the blue HB, for sources at radii $r > 19'$. Orange and red lines show the respective HB fiducials of M55 and $\omega$ Cen, which were constructed as described in Appendix B and have been individually scaled to distance moduli of $\mu_0^{M55} = 20.84$ and $\mu_0^{\omega\,Cen} = 20.78$ mag as described in Section 5. Dashed red lines show $2\sigma$ statistical uncertainties ($\pm 0.06$ mag) on the $\omega$ Cen fit. Right: The same as the left panel, for sources at radii $r < 19'$. Strong main sequence contamination of the blue HB is evident for stars at smaller radii.

To constrain a distance to Fornax from each cluster's calibrated ridgeline, we simply shift these calibrated distance moduli into alignment with the ridgeline defined by Fornax's blue HB. To enhance the contrast of the Fornax HB against contaminating stars (which are likely blue stragglers), we limit this comparison to radii $r > 19'$, as it is well established that Fornax's oldest and metal-poor populations tend to lie at larger radii (e.g., Battaglia et al. 2006; de Boer et al. 2012; Rusakov et al. 2021). The left panel of Figure 8 shows this outer population on the blue HB. The nearly coincident scaled GC ridgelines are plotted over the Fornax HB in orange and red, with dashed lines showing the $2\sigma$ statistical uncertainty of $\pm 0.06$ mag. Although defined and shifted without reference to one another, the two fiducials agree very well. In the right panel of Figure 8, the inner population at radii $r < 19'$ is plotted. A strong contaminating population of main sequence stars in this sample is clearly shown near $(V - I)_0 = 0.0$ mag.

The resulting distance moduli based on each of the clusters' ridgelines, fitting them individually by eye, are $\mu_0^{M55} = 20.84$ and $\mu_0^{\omega\,Cen} = 20.78$ mag (scaled in Figure 8). After accounting for their respective distances, we find that the two cluster ridgelines match Fornax's blue HB reasonably well, suggesting that so-called "second-parameter" effects in the HB are unlikely to be introducing inaccuracy in our distance estimations.

### 5.2. HB Distance and Error Budget

Based on the uncertainties associated with the DEB distances alone, the weighted-average distance modulus derived from this analysis is $\mu_0^{HB} = 20.83$ mag. We estimate the statistical uncertainty associated with this measurement to be half the difference between the distance moduli derived from the two individual clusters, i.e., 0.03 mag. As the two clusters cover a representative range in metallicity, averaging over both accounts for potentially significant uncertainties related to metallicity effects. We note that this new method provides an update to that cited in Freedman (2021b); the resulting distance modulus is 0.04 mag farther (2% in distance).

As shown in Figure 8, Fornax's blue HB at $r > 19'$ is only well defined for colors $(V - I)_0 > 0.1$ mag. In the companion paper by Q. H. Tran et al. (2022, in preparation), the color range used to compare HBs is found to have a significant effect on the scaled distance modulus of Sculptor. In particular, fitting the HB at $(V - I)_0 < 0.1$ mag separately from $(V - I)_0 > 0.1$ mag leads to a difference in derived distances of 0.05 mag. Because we can only clearly detect Fornax's HB in the $(V - I)_0 > 0.1$ mag regime, this effect may similarly introduce a systematic uncertainty at the same level. We therefore add an additional systematic term of 0.05 mag to account for a possible bias associated with our limited color range of observation.

Assuming systematics are shared between the two cluster calibrators, the cumulative M55 uncertainty can be taken as a systematic uncertainty floor for the DEB distances. We therefore adopt a systematic zero-point uncertainty of 0.05 mag, along with an uncertainty of 0.01 mag to account for possible variation in the photometric system. We note that this is a conservative estimate as all photometry is calibrated to the same Stetson–Landolt system. Based on the scatter of the HB-metallicity relation (Dotter et al. 2010; Federici et al. 2012), we adopt an additional systematic uncertainty of 0.04 mag. Finally, we include 15% of the foreground reddening (0.02 mag, Schlegel et al. 1998; Schlafly & Finkbeiner 2011) as the systematic uncertainty for the reddening to the calibrating clusters, as well as the 0.016 and 0.01 mag terms associated with our own reddening uncertainties.

The above uncertainties are compiled in Section 6.1 and Table 2. Combining statistical and systematic uncertainties in quadrature, we find a horizontal branch distance modulus to Fornax of $\mu_0^{HB} = 20.83 \pm 0.03_{stat} \pm 0.09_{sys}$ mag, in good agreement with the distances derived in this work from the TRGB and RR Lyrae.

## 6. Discussion

### 6.1. Systematic and Statistical Error Budgets

The comprehensive error budgets for each distance modulus are presented in Table 2.

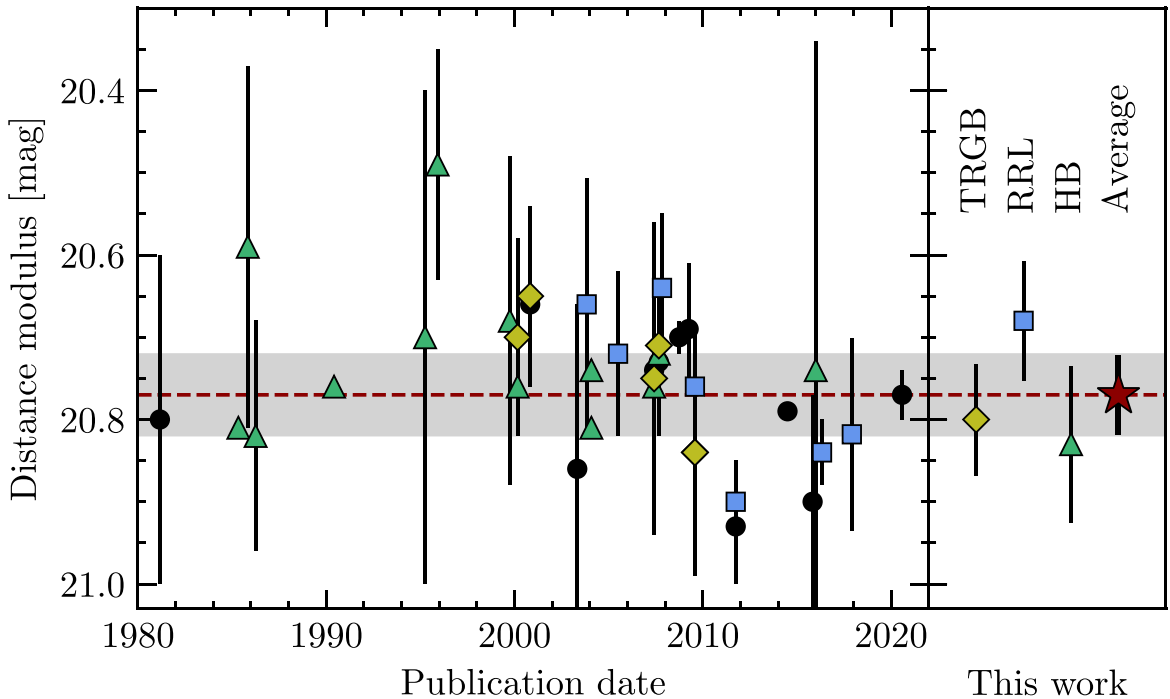

**Figure 9.** Comparison of literature distances to the Fornax dSph (for original data and references see Table 4). Distances based on the TRGB, RRL PL/PW relations or mean magnitudes, and the horizontal branch are shown as yellow diamonds, blue squares, and green triangles, respectively. Distances using other methods are shown in black. The dashed horizontal line at $\mu_0 = 20.770$ mag shows the final averaged distance modulus from this work, with the corresponding $\pm 1\sigma = 0.048$ confidence interval highlighted in gray. Literature reddening assumptions and zero-point calibrations have not been adjusted.

### 6.2. Literature Comparison

Fornax is a popular target for distance measurements, and the literature includes a large number of estimates based on observations of the TRGB, mean HB luminosity, red clump (RC) luminosity, $\delta$ Scuti variables, carbon stars (CS), and other calibrators. We compile these distance moduli in Appendix C (Table 4) and display them graphically with our results in Figure 9. Here, we compare literature measurements taking similar approaches to our own three methods and account for discrepancies where they arise. For each respective method, our distance modulus agrees well and is among the most precise values published.

A literature search returned six TRGB detections in Fornax, which we find agree well with our distance modulus $\mu_0^{\mathrm{TRGB}} = 20.80 \pm 0.037_{\mathrm{stat}} \pm 0.057_{\mathrm{sys}}$ mag. In the $K$–band, Gullieuszik et al. (2007) and Pietrzyński et al. (2009) report $\mu_0 = 20.75 \pm 0.19$ and $\mu_0 = 20.84 \pm 0.15$ mag, respectively. Pietrzyński et al. (2009) also find $\mu_0 = 20.84 \pm 0.12$ mag using the $J$–band TRGB. The three literature $I$-band detections all adopt a slightly higher zero-point calibration than this work. Adjusting their reddening and zero-point assumptions to match this paper (without altering their respective uncertainties) brings each into good agreement with our distance modulus: $\mu_0 = 20.74 \pm 0.12$ mag (Saviane et al. 2000), $\mu_0 = 20.70 \pm 0.11$ mag (Bersier 2000), and $\mu_0 = 20.77 \pm 0.07$ mag (Rizzi et al. 2007a).

Of the seven studies using RRLs as a distance indicator in Fornax, only one considers PLZ relations rather than taking a calibration based on the mean $V$ luminosity. Karczmarek et al. (2017) use NIR $JK$ observations of 77 BW02 RRLs and report a distance modulus $\mu_0 = 20.82 \pm 0.12$ mag. Although fainter than this work's distance modulus $\mu_0^{\mathrm{PWZ}} = 20.68 \pm 0.02_{\mathrm{stat}} \pm 0.07_{\mathrm{sys}}$ mag, the two measurements agree to within the reported uncertainties.

The Fornax literature additionally contains thirteen measurements using the HB as a primary distance indicator, of which nearly all are in agreement with our observed modulus $\mu_0^{\mathrm{HB}} = 20.83 \pm 0.03_{\mathrm{stat}} \pm 0.09_{\mathrm{sys}}$ mag. Twelve of these literature detections take the HB mean magnitude as their zero-point calibrator; Hendricks et al. (2016) is the only one which applies an approach more similar to our own HB fitting. They consider synthetic zero-age isochrones applied to Fornax's GC 4, finding a distance modulus $\mu_0 = 20.74 \pm 0.40$ mag in good agreement with this work. Our discrepancy with Irwin & Hatzidimitriou (1995)'s HB detection, $\mu_0 = 20.49 \pm 0.14$ mag, can entirely be explained by their choice of zero-point. Taking their parameters $V_{\mathrm{HB}} = 21.29$ mag and [Fe/H] $= -1.4$ dex, we can update their detection to $\mu_0 = 20.81$ mag using a more recent calibration from Cacciari & Clementini (2003), resolving any tension. We also note an apparent typo in their reported distance modulus (20.40 mag, rather than 20.49 mag), which we have additionally corrected for.

### 6.3. TRGB and JAGB Zero-Points

Using our RRL PWZ distance modulus $\mu_0^{\mathrm{RRL}} = 20.68 \pm 0.02_{\mathrm{stat}} \pm 0.07_{\mathrm{sys}}$ mag and the $I$-band TRGB detection $I_0 = 16.753 \pm 0.03_{\mathrm{stat}} \pm 0.037_{\mathrm{sys}}$ mag, we derive a TRGB zero-point calibration of $M_I^{\mathrm{RRL}}(\mathrm{TRGB}) = -3.93 \pm 0.04_{\mathrm{stat}} \pm 0.08_{\mathrm{sys}}$ mag. From the HB distance modulus $\mu_0^{\mathrm{HB}} = 20.83 \pm 0.03_{\mathrm{stat}} \pm 0.09_{\mathrm{sys}}$ mag we independently find $M_I^{\mathrm{HB}}(\mathrm{TRGB}) = -4.08 \pm 0.04_{\mathrm{stat}} \pm 0.10_{\mathrm{sys}}$ mag. These show good agreement with Freedman et al. (2020)'s TRGB zero-point of $M_I = -4.047 \pm 0.022_{\mathrm{stat}} \pm 0.039_{\mathrm{sys}}$ mag (updated to $M_I = -4.042$[7] $\pm\,0.015_{\text{stat}}\pm 0.035_{\text{sys}}$ in Freedman 2021a) and are consistent to within the errors. We note that our average $\langle M_I \rangle = -3.99 \pm 0.03_{\text{stat}} \pm 0.06_{\text{sys}}$ mag from RRL and HB calibrators agrees with the literature value at the 0.05 mag level.

---
[7] $M_{814} = -4.049$ mag.

Finally, we highlight the excellent agreement between this work and Freedman & Madore (2020)'s independently determined JAGB measurement of $\mu_0 = 20.77 \pm 0.03$ mag. Taking their reported JAGB peak at $J_0 = 14.57 \pm 0.03_{\text{stat}}$ mag, we find a TRGB-based zero-point of $M_J^{\text{TRGB}}(\text{JAGB}) = -6.23 \pm 0.05_{\text{stat}} \pm 0.06_{\text{sys}}$, an RRL-based zero-point of $M_J^{\text{RRL}}(\text{JAGB}) = -6.11 \pm 0.04_{\text{stat}} \pm 0.07_{\text{sys}}$, and an HB-based zero-point $M_J^{\text{HB}}(\text{JAGB}) = -6.26 \pm 0.04_{\text{stat}} \pm 0.09_{\text{sys}}$ mag. Their weighted average $\langle M_J \rangle = -6.19 \pm 0.03_{\text{stat}} \pm 0.04_{\text{sys}}$ mag agrees to 0.01 mag with the literature calibration $M_J(\text{JAGB}) = -6.20 \pm 0.037$ mag.

### 6.4. A Combined Distance to Fornax

This work has presented three independent distances, each anchored to different geometric zero-point calibrators and relying on unique stellar physics. The TRGB, RRL, and HB stellar populations are distinct as well, helping to mitigate cross-measurement systematics. We combine the TRGB, RRL PWZ, and HB distances as a product distribution for a cumulative distance modulus of $\langle \mu_0 \rangle = 20.770 \pm 0.042_{\text{stat}} \pm 0.024_{\text{sys}}$ mag. The 0.042 mag statistical uncertainty term is the width of this product distribution, excluding the shared photometric zero-point, extinction law, and line-of-sight reddening systematic error terms which cannot be averaged over. These contribute to form the 0.024 mag cumulative systematic uncertainty, combined as the root sum of squares. This distance modulus implies a true distance of $143 \pm 3$ kpc.

### 7. Conclusions

We have presented three high-precision, independent Population II measurements of the distance to the Fornax dSph using wide-field $V$ and $I$ imaging. From an edge-detection measurement of the TRGB apparent magnitude, we find $I_0^{\text{TRGB}} = 16.753 \pm 0.03_{\text{stat}} \pm 0.037_{\text{sys}}$ mag. Using the Freedman et al. (2020) TRGB zero-point calibration $M_I^{\text{TRGB}} = -4.047 \pm 0.022_{\text{stat}} \pm 0.039_{\text{sys}}$ mag, we determine a de-reddened distance modulus $\mu_0^{\text{TRGB}} = 20.80 \pm 0.037_{\text{stat}} \pm 0.057_{\text{sys}}$ mag (adopting $A_I = 0.032$ mag). For an RR Lyrae distance, we phase our IMACS photometry using periods from the archival catalogs of BW02 and MG03 and present updated period–luminosity–metallicity and period–Wesenheit–metallicity relations. Adopting Ne19's calibrations from Gaia DR2 parallaxes of galactic RR Lyraes, we determine separate distance moduli for PLZ and PWZ relations, taking $\mu_0^{\text{PWZ}} = 20.68 \pm 0.02_{\text{stat}} \pm 0.07_{\text{sys}}$ mag as the most representative. Finally, we find $\mu_0^{\text{HB}} = 20.83 \pm 0.03_{\text{stat}} \pm 0.09_{\text{sys}}$ mag using Fornax's blue horizontal branch and a zero-point calibration based on DEB distances to two galactic globular clusters.

These distances are in excellent agreement with previous determinations and demonstrate encouraging convergence and consistency between the three methods. They improve upon the existing literature by providing very precise measurements for their respective methods, with a thorough treatment of metallicity and population effects. As they derive from homogeneous photometry, are anchored to independent geometric zero-points, and utilize different classes of Population II stars, we adopt a combined distance modulus of $\langle \mu_0 \rangle = 20.770 \pm 0.042_{\text{stat}} \pm 0.024_{\text{sys}}$ mag, corresponding to a distance of $143 \pm 3$ kpc.

We thank Dylan Hatt for refining our analysis procedures and for helpful discussions. Thanks as well to Peter Stetson for his advice and assistance troubleshooting DAOPHOT photometry. We also thank the Observatories of the Carnegie Institution for Science and the University of Chicago for their support of long-term research into the calibration and determination of the expansion rate of the Universe. We are grateful to the anonymous referee for helpful comments regarding the manuscript.

This research has made use of the NASA/IPAC Extragalactic Database (NED), which is operated by the Jet Propulsion Laboratory, California Institute of Technology, under contract with the National Aeronautics and Space Administration.

Based, in part, on observations taken with the IMACS camera on Baade/Magellan 6.5 m telescope at the Las Campanas Observatories, which is owned and operated by the Carnegie Institution for Science.

Some of the data presented in this paper were obtained from the Mikulski Archive for Space Telescopes (MAST).

Support for program #13691 was provided by NASA through a grant from the Space Telescope Science Institute, which is operated by the Association of Universities for Research in Astronomy, Inc., under NASA contract 5-26555.

*Facilities:* HST (ACS/WFC), Magellan-Baade (IMACS).

*Software:* DAOPHOT (Stetson 1987), ALLFRAME (Stetson 1994), TinyTim (Krist et al. 2011), Astropy (Astropy Collaboration et al. 2013, 2018), SciPy (Virtanen et al. 2020), TOPCAT (Taylor 2005).

### Appendix A
### Fornax IMACS Catalog

Table 3 presents the header and first four lines of data for the Fornax $V, I$ photometric catalog. Sources were reduced from 68 mosaic IMACS images and calibrated to CADC standards as in Section 2.2. For overlapping sources across redundant images, RA, Dec, $V$-band flux, and $I$-band flux were averaged. Fluxes were converted back to magnitudes ($V\text{mag}$ and $i\text{mag}$). Magnitude errors ($e_{V\text{mag}}$ and $e_{i\text{mag}}$) were estimated as a sum of squares with DAOPHOT magnitude errors and the standard deviation of the calibrating image (e.g., Figure 2 panels (c) and (d)), and should not be interpreted as rigorous estimates of the photometric error. Sources excluded by Gaia EDR3 proper motions are flagged with "1" in the CutFlag column. The full catalog contains 180,471 sources.

### Appendix B
### Globular Cluster Horizontal Branch Fiducials

Figure 10 shows the cluster photometry for M55 and $\omega$ Cen which was used to calibrate our HB distance moduli for Fornax (this work) and Sculptor (Q. H. Tran et al. 2022, in preparation). As described in Section 5.1, the ridgeline fiducials (orange and red lines) are defined based on Stetson et al. (2019) cluster photometry cleaned using Gaia EDR3 proper motions. We construct empirical ridgelines by sampling the three-sigma-clipped median $I$-band magnitude of the blue HB in color intervals of 0.025 mag, with the fiducial HB line computed

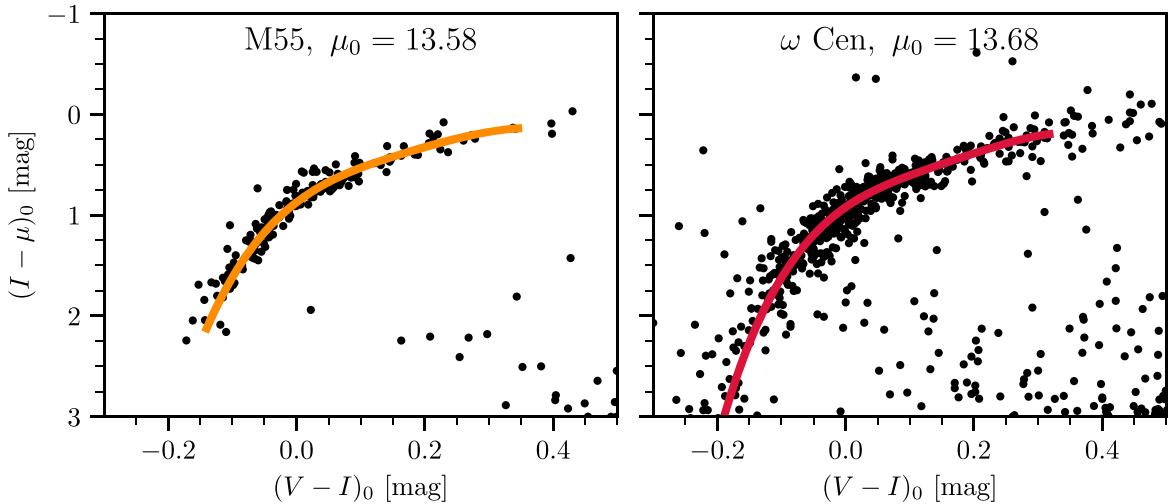

**Figure 10.** HB ridgeline fiducials (orange and red lines) over Stetson et al. (2019) homogeneous cluster photometry (black points) for M55 and $\omega$ Cen. Fiducial ridgelines are empirically defined by sampling the median magnitude of each cluster's HB in intervals of 0.025 mag along the color axis and fitting a smoothed spline to these points, calibrated using distances and reddenings from the literature.

**Table 3**
Header and Sample Data for IMACS Photometry ($N = 180{,}471$)

| RAdeg [deg] | DEdeg [deg] | Vmag [mag] | $e_{Vmag}$ [mag] | imag [mag] | $e_{imag}$ [mag] | CutFlag [bool] |
|---|---|---|---|---|---|---|
| 39.7673182 | −34.7253171 | 20.842 | 0.0488 | 20.532 | 0.0291 | 0 |
| 39.7505466 | −34.2924040 | 23.900 | 0.0688 | 23.217 | 0.0703 | 0 |
| 39.9946150 | −34.7811629 | 21.383 | 0.0416 | 20.570 | 0.0358 | 0 |
| 40.3163033 | −34.7638341 | 18.284 | 0.0286 | 16.905 | 0.0262 | 0 |

(This table is available in its entirety in machine-readable form.)

from a spline interpolation. To scale the respective photometry and calibrate the resulting HB curves of each cluster as in Figure 10, we adopt distances and reddening values from the literature. We take $\mu_0 = 13.58$ mag and $E(B-V) = 0.115$ for M55 (Kaluzny et al. 2014), and $\mu_0 = 13.68$ mag and $E(B-V) = 0.12$ for $\omega$ Cen (Thompson et al. 2001).

## Appendix C
## Literature Distances to Fornax

See Table 4 for a comprehensive compilation of literature distance moduli to Fornax. Where statistical and systematic uncertainties were differentiated, they have been added in quadrature. Otherwise, quoted values are unchanged from the literature except where indicated by a footnote.

## Appendix D
## Relative Calibration of HST Photometry

Accurate ground-to-HST transformations are essential to measuring $H_0$ via geometric anchors in the Local Group, i.e., for CCHP observations of SNe Ia host galaxies (Freedman et al. 2019; Hoyt et al. 2021). The $I$-band to ACS/WFC F814W transformation in particular is a significant source of systematic uncertainty in the TRGB zero-point calibration (Beaton et al. 2016; Freedman et al. 2019; Jang et al. 2021). In nearby galaxies such as Sculptor and Fornax, wide-field, ground-based observations can provide strong statistical sampling of RGB stars to complement deep HST photometry.

To determine a ground-to-HST calibration, we matched overlapping sources between the ground-based IMACS $V$- and $I$-band photometry and HST ACS/WFC F606W and F814W photometry, the footprints of which we have compared in Figure 1. We first applied magnitude limits of 21.5 and 15.0 mag and a 1 mag photometric uncertainty limit to the HST and Stetson-calibrated IMACS catalogs described in Section 2. This was done to exclude saturated and poorly resolved sources, as well as to ensure that our catalog is primarily dominated by an RGB population. We matched sources with centers within 1″.8 of each other and visually reviewed each match to remove those that appeared crowded, non-stellar, or on the edge of a CCD, leaving a sample of 220 pairs.

Panels (a) and (b) of Figure 11 show the color transformations between these calibrated ground-based $V$ and $I$ magnitudes and the corresponding F606W and F814W flight magnitudes. Sources greater than three standard deviations from a linear fit to either distribution were excluded, shown in gray (cut from the respective panel's distribution) and black (if cut from another panel's distribution). For the $I - $F814W relation of panel (b), a dashed horizontal line shows the mean value of $-0.002 \pm 0.038$ mag, where the uncertainty is taken to be the standard deviation of the residuals. This value is consistent with zero, in agreement with Q. H. Tran et al. (2022, in preparation)'s $I - $F814W calibration in Sculptor. Our calibration is also plotted against the $I$-band magnitude in panel (d) with the same $3\sigma$ cut.

For the color-dependent $V - $F606W transformation in panel (a), the dashed line shows a linear fit with $\pm 1\sigma$ regions in gray. We find a slope of $0.303 \pm 0.023$ mag/mag that is consistent with the value $0.236 \pm 0.058$ mag/mag found by Sirianni et al. (2005). At the mean color $\langle (V-I) \rangle = 0.96$ mag, the ZP

Table 4
Literature Distance Moduli to the Fornax dSph

| $\mu_0$ [mag] | Method | Filter | Zero-Point | Reference |
|---|---|---|---|---|
| 20.8 ± 0.2 | RGB fit | B,V | Sandage (1970) | Verner et al. (1981) |
| 20.81 | HB | V | Harris (1980) | Webbink (1985) |
| 20.59 ± 0.22 | HB | V | $M_V = 0.6$ mag | Buonanno et al. (1985) |
| 20.82 ± 0.14 | HB | V | $M_V = 0.6$ mag | Gratton et al. (1986) |
| 20.76 | HB | V | $M_V = 0.6$ mag | Demers et al. (1990) |
| 20.7 ± 0.3 | HB | V | $M_V = 0.6$ mag | Beauchamp et al. (1995) |
| 20.49 ± 0.14 | HB | V | Carney et al. (1992) | Irwin & Hatzidimitriou (1995) [a] |
| 20.68 ± 0.20 | HB | V | Lee et al. (1990) | Buonanno et al. (1999) |
| 20.70 ± 0.12 | TRGB | I | Da Costa & Armandroff (1990) | Saviane et al. (2000) |
| 20.76 ± 0.04 | HB | V | Lee et al. (1990) | Saviane et al. (2000) |
| 20.66 | RC | I | Popowski (2000) | Bersier (2000) |
| 20.65 ± 0.11 | TRGB | I | Lee et al. (1993) | Bersier (2000) |
| 20.86 ± 0.20 | RC | K | Alves (2000) | Pietrzyński et al. (2003) [b] |
| 20.66 ± 0.15 | ⟨RRL⟩ | V | Chaboyer (1999) | Mackey & Gilmore (2003) |
| 20.74 | HB | I | Demarque et al. (2000) | Pont et al. (2004) |
| 20.81 | HB | V | Demarque et al. (2000) | Pont et al. (2004) |
| 20.72 ± 0.10 | ⟨RRL⟩ | V | Gratton et al. (2003) | Greco et al. (2005) |
| 20.74 ± 0.11 | RC | K | Alves (2000) | Gullieuszik et al. (2007) |
| 20.75 ± 0.19 | TRGB | K | Valenti et al. (2004) | Gullieuszik et al. (2007) |
| 20.71 ± 0.07 | TRGB | I | Bellazzini et al. (2001) | Rizzi et al. (2007a) |
| 20.72 ± 0.06 | HB | V | Cacciari & Clementini (2003) | Rizzi et al. (2007a) |
| 20.73 ± 0.09 | RC | I | Stanek et al. (1998) | Rizzi et al. (2007a) |
| 20.76 ± 0.04 | HB | V | Caretta (2000) | Rizzi et al. (2007b) |
| 20.64 ± 0.09 | ⟨RRL⟩ | V | Cacciari & Clementini (2003) | Greco et al. (2007) [c] |
| 20.70 ± 0.02 | SXP PLZ | V | McNamara et al. (2004) | Poretti et al. (2008) |
| 20.69 ± 0.08 | Mira PL | K | van Leeuwen et al. (2007) | Whitelock et al. (2009) |
| 20.76 ± 0.07 | ⟨RRL⟩ | V | Cacciari & Clementini (2003) | Greco et al. (2009) [d] |
| 20.84 ± 0.12 | TRGB | J | Valenti et al. (2004) | Pietrzyński et al. (2009) |
| 20.84 ± 0.15 | TRGB | K | Valenti et al. (2004) | Pietrzyński et al. (2009) |
| 20.93 ± 0.07 | δS & SXP PLZ | V | Derived PLs | McNamara (2011) |
| 20.90 ± 0.05 | ⟨RRL⟩ | V | $M_V = 0.53, 0.34$ mag | McNamara (2011) |
| 20.79 | CMD fit | V,I | Padova isochrones | Weisz et al. (2014) |
| 20.90 ± 0.13 | CS PL | $W_{K,J-K}$ | Skrutskie et al. (2006) | Huxor & Grebel (2015) [e] |
| 20.74 ± 0.40 | HB fit | V | Dotter et al. (2008) | Hendricks et al. (2016) |
| 20.84 ± 0.04 | ⟨RRL⟩ | V | Cacciari & Clementini (2003) | de Boer & Fraser (2016) |
| 20.82 ± 0.12 | RRL PLZ | J,K | Sollima et al. (2008) | Karczmarek et al. (2017) |
| 20.77 ± 0.03 | JAGB | J | Pietrzynski et al. (2019) | Freedman & Madore (2020) |
| 20.80 ± 0.07 | TRGB | I | Freedman et al. (2020) | This work |
| 20.68 ± 0.07 | RRL PWZ | $W_{I,V-I}$ | Neeley et al. (2019) | This work |
| 20.83 ± 0.09 | HB fit | V,I | Thompson et al. (2001); Kaluzny et al. (2014) | This work |
| 20.77 ± 0.05 | Combined | … | … | This work |

**Notes.**
[a] Accounting for an apparent typo in the reported distance modulus (20.40 rather than 20.49 mag).
[b] Additional systematic error of 0.02 mag, per discussion of photometric calibration.
[c] Adopting the low-metallicity estimate.
[d] Adopting the low-metallicity estimate.
[e] Average over 18 carbon stars.

offset is $0.210 \pm 0.004$ mag. Defining $V'$ as the $V$ magnitude corrected for this color term, we present a calibration of $V'$ to F606W against magnitude in panel (c).

Smoothed kernel density estimations (KDEs) for the respective $V'-$ F606W and $I-$ F814W distributions of panels (c) and (d) are also shown. These exhibit symmetrical Gaussian distributions about their mean values, which are highly consistent with a transformation of zero in both cases. The standard deviation $\sigma = 0.04$ mag for both fits, demonstrating that the color term $V'$ with a slope of 0.303 mag/mag effectively corrects for the functionality seen in the $V-$ F606W transformation.

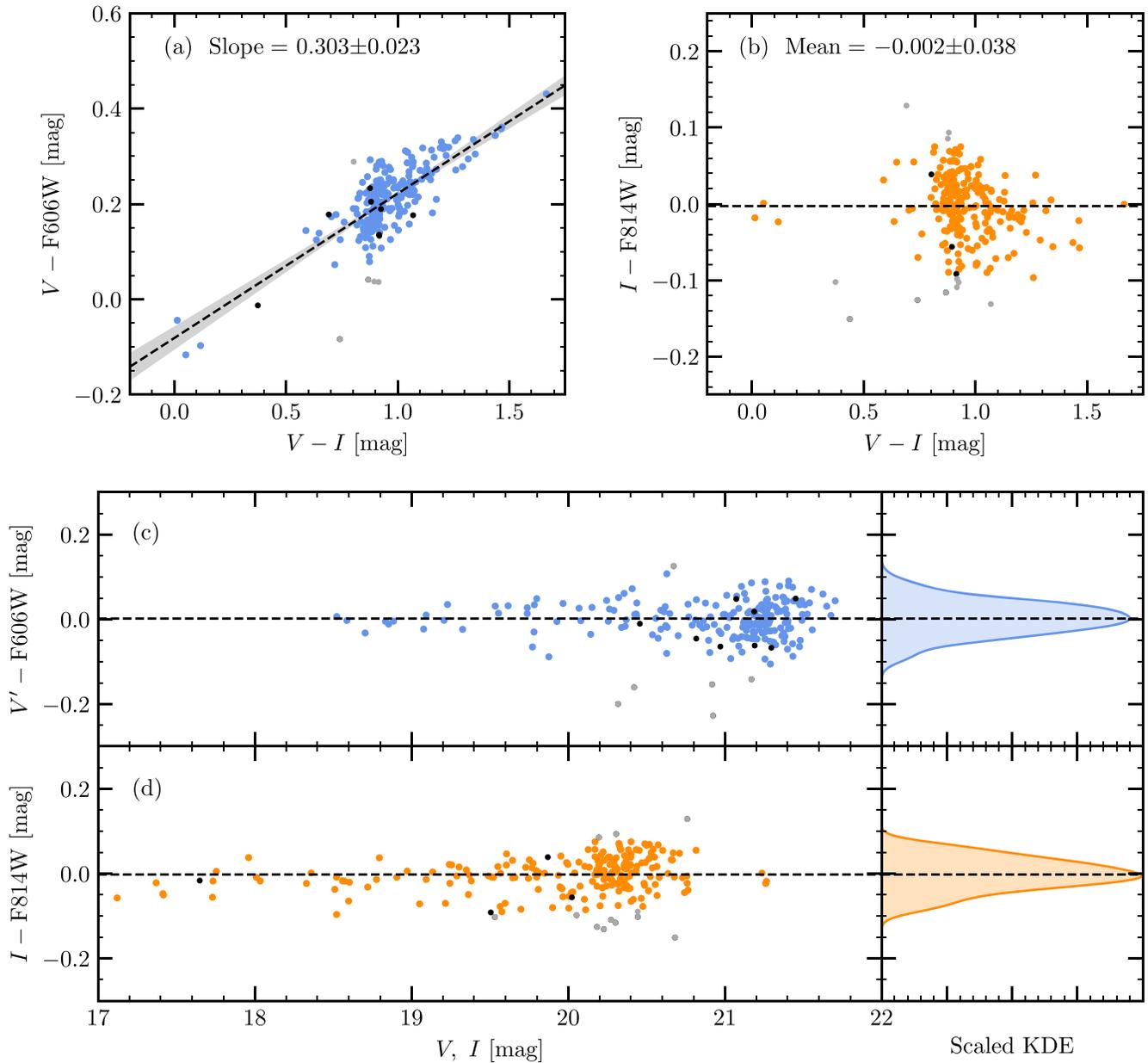

**Figure 11.** Ground-to-HST transformations for 220 RGB stars matched between CCHP Stetson-calibrated IMACS data and ACS/WFC photometry (Section 2), cleaned and matched as described in Appendix D: (a) Stetson $V$ to F606W against Stetson $(V-I)$ color; (b) Stetson $I$ to F814W against Stetson $(V-I)$ color; (c) $V'$ to F606W against $V$ magnitude, where $V'$ is the $V$ magnitude corrected for the linear color term of panel (a); (d) Stetson $I$ to F814W against $I$ magnitude. In each panel, outlier data more than $3\sigma$ from a linear fit are excluded and plotted in gray. Data failing this cut in either color transformation are additionally excluded and plotted in black for all panels. For panels (b), (c), and (d), black dashed lines show mean differences consistent with zero, while for the $V$ to F606W transformation of panel (a), the black dashed line shows a linear fit with a slope of 0.303 mag/mag and $\pm 1\sigma$ regions in gray.

**ORCID iDs**

Elias K. Oakes ● https://orcid.org/0000-0002-0119-1115
Taylor J. Hoyt ● https://orcid.org/0000-0001-9664-0560
Wendy L. Freedman ● https://orcid.org/0000-0003-3431-9135
Barry F. Madore ● https://orcid.org/0000-0002-1576-1676
Quang H. Tran ● https://orcid.org/0000-0001-6532-6755
William Cerny ● https://orcid.org/0000-0003-1697-7062
Rachael L. Beaton ● https://orcid.org/0000-0002-1691-8217
Mark Seibert ● https://orcid.org/0000-0002-1143-5515